\documentclass[iicol, sn-nature]{sn-jnl}

\usepackage{graphicx}%
\usepackage{multirow}%
\usepackage{amsmath,amssymb,amsfonts}%
\usepackage{amsthm}%
\usepackage{mathrsfs}%
\usepackage[title]{appendix}%
\usepackage{xcolor}%
\usepackage{textcomp}%
\usepackage{manyfoot}%
\usepackage{booktabs}%
\usepackage{algorithm}%
\usepackage{algorithmicx}%
\usepackage{algpseudocode}%
\usepackage{listings}%

\begin{document}

\title{Machine Learning Based Compton Suppression for Nuclear Fusion Plasma Diagnostics}

\author*[1,2]{\fnm{Kimberley} \sur{Lennon}}\email{kimberley.lennon@ukaea.uk}

\author[2]{\fnm{Chantal} \sur{Shand}}\email{chantal.shand@ukaea.uk}

\author[1]{\fnm{Robin} \sur{Smith}}\email{robin.smith@shu.ac.uk}

\affil[1]{\orgdiv{Materials and Engineering Research Institute}, \orgname{Sheffield Hallam University}, \orgaddress{\street{Howard Street}, \city{Sheffield}, \postcode{S1 1WB}, \country{UK}}}

\affil[2]{\orgdiv{Applied Radiation Technology Group}, \orgname{UK Atomic Energy Authority}, \orgaddress{\street{Culham Campus}, \city{Abingdon}, \postcode{OX14 3DB}, \country{UK}}}


\abstract{Diagnostics are critical on the path to commercial fusion reactors, since measurements and characterisation of the plasma is important for sustaining fusion reactions. Gamma spectroscopy is commonly used to provide information about the neutron energy spectrum from activation analysis, which can be used to calculate the neutron flux and fusion power. The detection limits for measuring nuclear dosimetry reactions used in such diagnostics are fundamentally related to Compton scattering events making up a background continuum in measured spectra. This background lies in the same energy region as peaks from low-energy gamma rays, leading to detection and characterisation limitations. This paper presents a digital machine learning Compton suppression algorithm (MLCSA), that uses state-of-the-art machine learning techniques to perform pulse shape discrimination (PSD) for high purity germanium (HPGe) detectors. The MLCSA identifies key features of individual pulses to differentiate between those that are generated from photopeaks and Compton scatter events. Compton events are then rejected, reducing the low energy background. This novel suppression algorithm improves gamma spectroscopy results by lowering minimum detectable activity (MDA) limits and thus reducing the measurement time required to reach the desired detection limit. In this paper, the performance of the MLCSA is demonstrated using an HPGe detector, with a gamma spectrum containing americium-241 (Am-241) and cobalt-60 (Co-60). The MDA of Am-241 improved by 51\% and the signal to noise ratio (SNR) improved by 49\%, while the Co-60 peaks were partially preserved (reduced by 78\%). The MLCSA requires no modelling of the specific detector and so has the potential to be detector agnostic, meaning the technique could be applied to a variety of detector types and applications.}


\maketitle

\section{Introduction}\label{sec:intro}

\begin{figure}[h]
    \centering
    \includegraphics[width=0.4\textwidth]{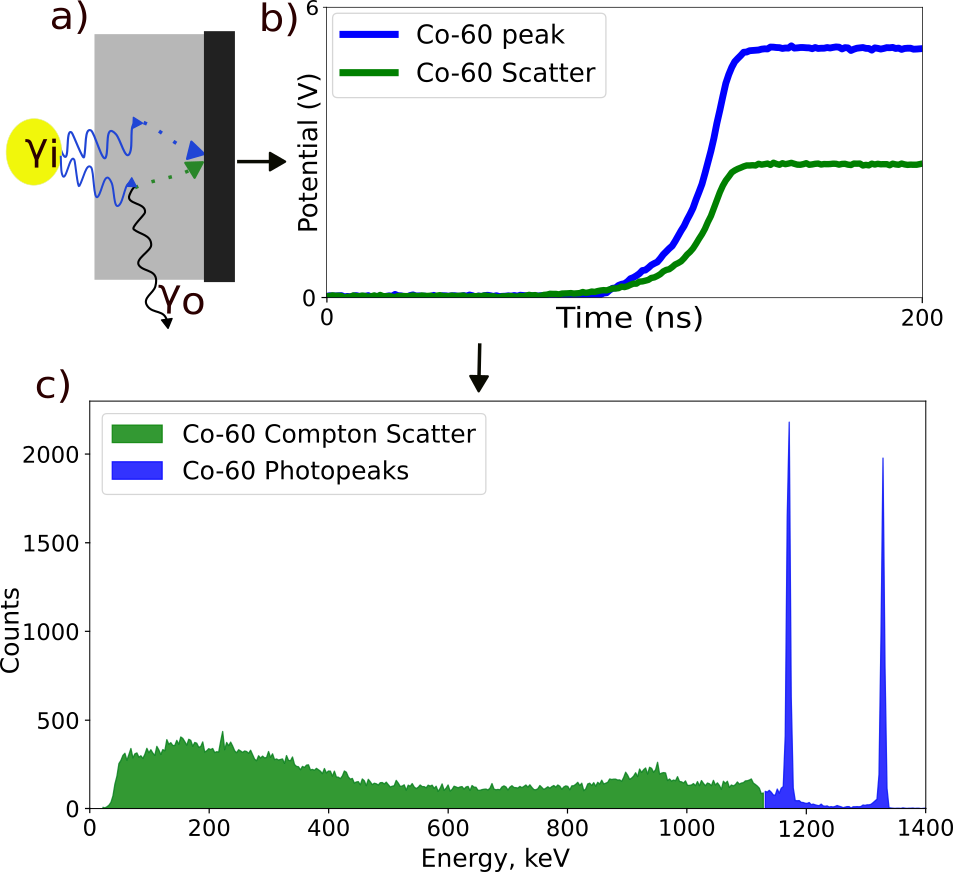}
    \caption{a) Gamma-ray interactions in a HPGe crystal: one is fully photoelectrically absorbed and produces a full energy photopeak (blue/dark) and the other Compton scatters and contributes to the Compton continuum (green/light). b) Example pulses from each interaction type. c) A typical energy spectrum showing Compton and photoelectric events}
    \label{fig:Compton_scatter}
\end{figure}

Gamma spectroscopy is a common method used in the nuclear industry to identify and quantify the presence of radiation. High purity germanium (HPGe) detectors are often selected to measure low intensity or complex gamma-ray signatures due to their excellent $\sim$~keV resolution and will be the focus of this paper. In the nuclear fusion field, gamma spectroscopy is used in waste characterisation, materials research for future fusion machines, neutron flux quantification via activation foils \cite{Lees_paper}, and many other areas. The nuclear structure of radionuclides common to fusion research mean it is often the case that low activity nuclides emitting low energy $\gamma$ rays need to be identified in the presence of higher energy $\gamma$ emitters. For example, the neutron activation of composite materials, such as stainless steel, readily produces complex activation networks, resulting in the emission of a vast spectrum of $\gamma$ energies \cite{GoodellJ.}. This poses a problem, as when higher energy photons interact within the detector they can Compton scatter \cite{krane, practical_g_spec} out of the crystal and only deposit a fraction of their energy. This contributes to a continuum of detected energies, which extends to the lower energy part of the spectrum (figure~\ref{fig:Compton_scatter}). For example, in irradiated steel, the presence of Co-57 is indicated by a low-energy 122~keV gamma ray, which will be masked by the Compton scattering contribution of high energy Co-60 $\gamma$ rays of 1173 and 1332~keV. This elevated background increases the low energy minimum detectable activities (MDA), which negatively impacts the characterisation work.

\begin{figure}[h]
    \centering
    \includegraphics[width=0.4\textwidth]{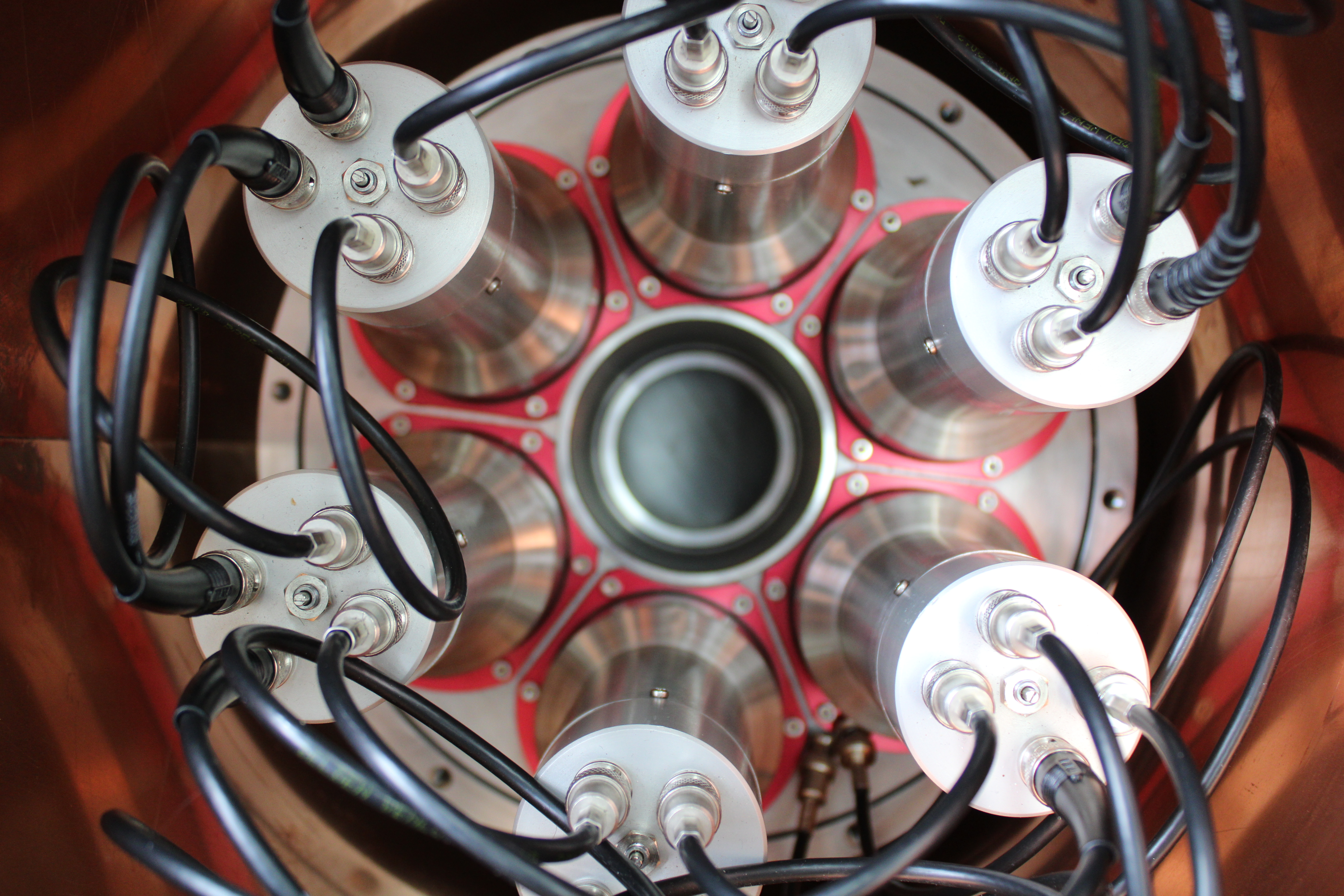}
    \caption{The Mirion broad energy germanium (BEGe) detector used in this work, with the physical Compton suppression system surrounding the main HPGe crystal}
    \label{fig:BEGe_CSS_ring}
\end{figure}

There are some existing solutions to reduce the Compton scatter influence in gamma-ray spectra. One example is a Compton veto ring, where a ring of gamma spectroscopy detectors, typically sodium iodide (NaI), surround a HPGe crystal. An example of such a system is shown in figure~\ref{fig:BEGe_CSS_ring}, located at the radiological assay and detection lab (RADLab), at the United Kingdom atomic energy authority (UKAEA) in Oxfordshire. If a photon is detected in the HPGe crystal and NaI crystal within a set time window, the signal is rejected from the spectrum as a Compton scatter event. This method is effective at reducing the Compton continuum, but it often falsely reduces the photopeaks of interest \cite{callums_css}. A physical Compton suppression system is also bulky, expensive, and is not easily re-configurable for multiple detector types.

In this work, the physical Compton suppression system used at UKAEA has been demonstrated to reduce the Compton background on an Am-241 (59~keV photopeak) and Co-60 spectrum by $\sim$~ 88\%, however it reduced the Co-60 photopeaks by $\sim$~ 77\% and reduced the Am-241 photopeak by 18\% (based on equation \ref{eq5} in section \ref{evaluation_metrics}, peak height reduction as a percentage for energy regions of interest, discussed in sections \ref{evaluation_metrics} to \ref{eval_performance}). 

An alternative technique is to use digital algorithms that do not rely on additional hardware like a Compton veto ring. Past attempts \cite{Tree2019} have utilised a pulse rise time cut off to preserve low energy pulses corresponding to photoelectric absorption, while removing pulses contributing to the Compton continuum. Low energy photons are more likely to undergo photoelectric absorption close to the surface of the detector, resulting in a long charge collection time in the HPGe as the electrons drift further through the crystal to the collection electrodes. In contrast, high energy photons more readily pass through the detector, and are more likely to Compton scatter and deposit their energy further inside the crystal. This results in a shorter rise time of the generated electrical pulses. Previous solutions have rejected all pulses below a rise time threshold \cite{Tree2019} (therefore removing pulses from high energy photons) and required extensive modelling of the specific detector to determine the optimal rise time cutoff. While effective, it relies on classification based on a single parameter, is not practical for preserving a broad range of energies, and is difficult to deploy to multiple types of detectors as the models would require modification.

Another digital technique is the use of machine learning algorithms for pulse shape discrimination (PSD). Examples of this include gamma particle tracking \cite{Holloway} and $\alpha$-$\gamma$ PSD \cite{gamma_alpha_psd}. However $\gamma$-$\gamma$ PSD for Compton suppression on HPGe detectors has not been undertaken. This work presents a novel digital solution to Compton suppression, the machine learning Compton suppression algorithm (MLCSA), which performs Compton suppression (via $\gamma$-$\gamma$ PSD) on pre-amplified pulses to reduce the Compton continuum, while preserving higher energy photopeaks and not relying on any detector modelling or additional hardware. The MLCSA comprises a convolutional neural network (CNN) as a supervised, classification machine learning model. Incoming pulses are processed and the CNN classifies each pulse as either a pulse from a photoelectric absorption (photopeak) or from a Compton scatter event (background). The scattered pulses are rejected and the output is a spectrum containing only pulses classified as photopeaks.

\section{Methods}\label{sec2}
\begin{figure}[h]
    \centering
    \includegraphics[width=0.1\textwidth]{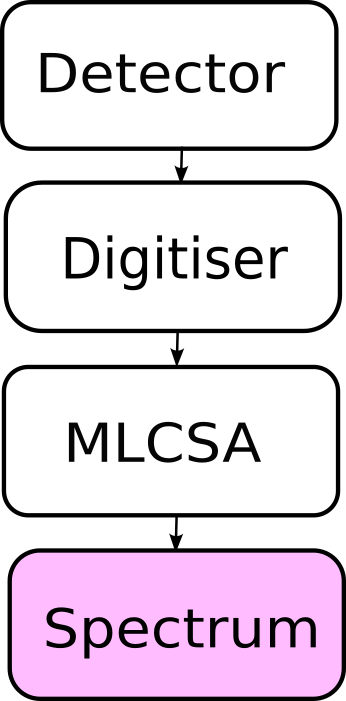}
    \caption{Digital Compton suppression flow diagram showing the four components, including the MLCSA}
    \label{fig:MLCSA}
\end{figure}

The MLCSA is an algorithm that performs Compton suppression on pulses from a HPGe detector and is part of a process that comprises four components, as shown in figure~\ref{fig:MLCSA}: detector (data acquisition), digitiser (signal/pulse processing), MLCSA (classifying), and Compton-suppressed spectrum (results). The methodology to these components are described in the following sections. 

\subsection{Detector and data collection}
The HPGe detector used to gather training data (which are later split into training and testing data) and live data (live data refers to an unseen, `realistic' data set, one which might be seen in a standard lab measurement of a mixed $\gamma$ source) was the Mirion broad energy germanium (BEGe) detector \cite{2021BEGeDetectors} (figure~\ref{fig:BEGe_CSS_ring}, model: BE3825, serial number: b13135). Five calibration sources were measured using the BEGe detector in this work to obtain training data: americium-241 (Am-241), cobalt-60 (Co-60), barium-133 (Ba-133), caesium-137 (Cs-137), and manganese-54 (Mn-54). All sources except for the Mn-54 source were positioned 10~cm from the end cap for measuring to reduce coincidence summing effects \cite{krane}. The Mn-54 source was placed at 0 cm due to its lower activity. For collection of the training data, each source was placed on the detector individually. Based on the energies of the detected pulses, these training data were split into \emph{Compton} and \emph{photopeak} categories. For the live data scenario, both the Am-241 and Co-60 sources were placed near the detector simultaneously (10~cm from the end cap) to create a realistic multi-nuclide spectrum.

\subsection{Digital pulse processing} \label{sec:dpp}
The pre-amplified pulses, referred to as raw pulses, were collected and processed by a Red Pitaya digitiser (the STEMlab 125-14 \cite{redpitaya}) with the trigger settings set as, Voltage range: (0, 20) V; Voltage trigger:  (-0.35, -0.4) V; trigger edge: positive; length of pulse: 200~ns.

\begin{figure}[h]
    \centering
    \includegraphics[width=0.1\textwidth]{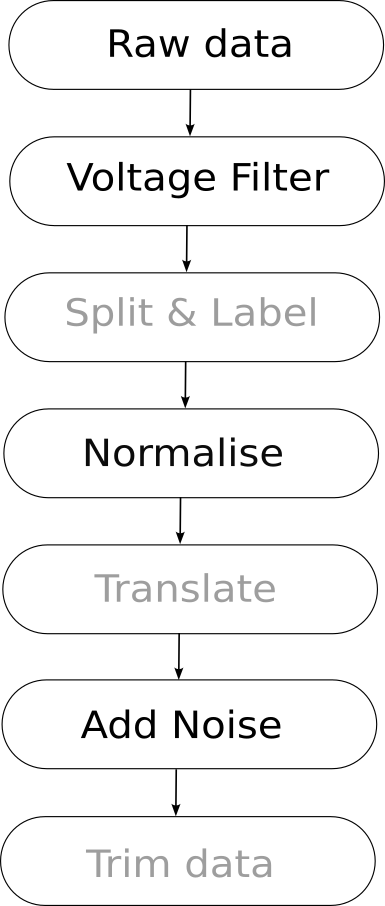}
    \caption{Pulse processing pipeline which is described in detail in the main text. The greyed out steps are not used when applying this process to the test and live data set}
    \label{fig:ml_pipeline}
\end{figure}

The measured raw pulses were subsequently modified so that the CNN could be trained purely on the shape of the pulses. Raw pulses could not be used directly for training the CNN as the algorithm was found to use features such as pulse height and noise level to characterise pulses in a way that lead to overfitting \cite{ML_handson_book}. Therefore, a pulse processing pipeline, shown in figure~\ref{fig:ml_pipeline}, was created to process the pulses for the training algorithm. This process included filtering, labelling, transformations, and regularisation techniques \cite{ML_handson_book} to reduce overfitting.

\begin{figure}[h]
    \centering
    \includegraphics[width=0.4\textwidth]{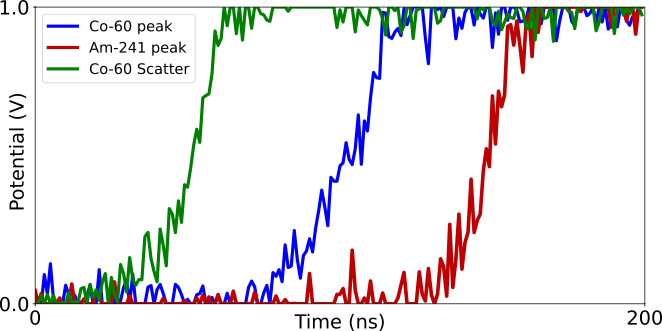}
    \caption{Examples of the final training pulses, with a Co-60 photopeak pulse in blue, a Co-60 Compton scatter pulse in green, and an Am-241 pulse in red}
    \label{fig:final_pulse}
\end{figure}

The pipeline starts by importing raw data, and removing pulses that exceed a maximum voltage and those below a pre-determined noise level. Then the data for each individual source are split into photopeak and scatter categories and labelled accordingly (based on their pulse heights). Pulses are then normalised to remove pulse height information and are randomly translated along the time axis. Gaussian noise is then added so that every pulse has the same noise level, irrespective of its energy. Finally the number of events in the photopeak and scatter categories are standardised. Example pulses for Am-241 and Co-60 photopeaks, and Co-60 Compton scatter, are shown in figure~\ref{fig:final_pulse}.

Once the complete training data set (consisting of five nuclides) had been collected and processed through the pipeline, it was split into two sub sets, as is common in machine learning training: a training set (80\%) and a testing set (20\%).

The processing pipeline for live data was similar to the training pipeline, except for the removal of the greyed out steps in figure~\ref{fig:ml_pipeline}. The split \& label step was removed because that information cannot be known for a realistic mixed $\gamma$ source. The time translation step was removed because it is unnecessary due to the way the CNN was trained. The trim data step was also not possible because the data originate from a mixed source hence the number of each type of event present is unknown.

\subsection{MLCSA}
\subsubsection{Convolutional Neural Networks}
The MLCSA is a supervised classification machine learning model, and specifically uses a convolutional neural network (CNN). The CNNs are a class of artificial neural networks and are tailored for handling structured grid-like data. They are typically used in fields such as object recognition, image classification, and image segmentation. They comprise multiple layers, including convolutional and pooling layers, and excel at feature extraction and pattern recognition within images  \cite{alexnet_cnn}. The convolutional layers use learnable filters to scan and identify local features, while pooling layers reduce spatial dimensions, enhancing computational efficiency and maintaining feature robustness \cite{ML_handson_book}.

The CNN architecture used in this work is a basic one-dimensional (1D) CNN \cite{1D_CNN}, specifically a sequential model with convolutional, max pooling, flatten, and dense layers (convolutional and pooling layers for feature extraction and dense layers for classification). This architecture is suitable for processing 1D sequences, such as time series data or any other 1D data \cite{ML_handson_book}. The architecture was implemented through standard \emph{sklearn} and \emph{keras} Python libraries.

\subsubsection{Training and Testing process}
\begin{figure}[h]
    \centering
    \includegraphics[width=0.3\textwidth]{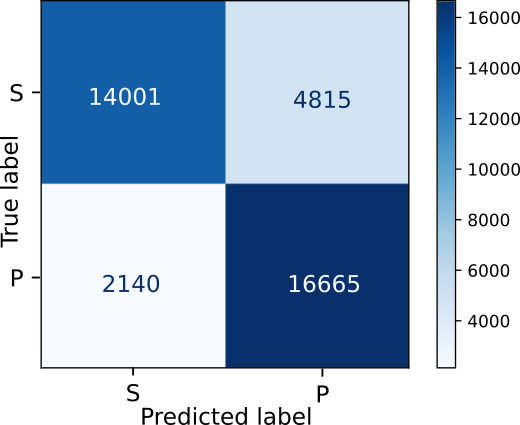}
    \caption{Confusion matrix for the pulses, corresponding to the optimal model parameters during training, where S is scatter and P is photopeak. The rows represent the true label and the columns represent the predicted labels}
    \label{fig:CNN_confusion_matrixl}
\end{figure}

The parameters (number of hidden layers, batches, and epochs) for the CNN in the MLCSA were selected and tuned using a process of random search hyper-parameter optimisation to find the best values for each \cite{ML_handson_book}. Each set of parameters were evaluated by checking the effect on the model performance, using common methods such as a confusion matrix, accuracy, precision, recall, and cross-validation (CV) score \cite{ML_handson_book, bishop_ml_book}. These evaluation methods all compare the known labels to the model predicted labels and all except the confusion matrix are recorded as a percentage. This process was repeated until there were no further improvements to the evaluation results (results didn't increase in value), and at that point the CNN was concluded to have reached peak performance. The final parameters that provided the best results (shown in section \ref{trained_cnn}), and therefore the ones used in the final CNN model were: 4 layers (3 hidden), 128 batches, and 30 epochs. The confusion matrix for the final model is shown in figure~\ref{fig:CNN_confusion_matrixl}, where a perfect model would have none-zero values only on its main diagonal (top left to bottom right) \cite{ML_handson_book}.

Once the CNN parameters were tuned, the model was trained using the 80\% split part of the full training data set, which had been processed as described in the steps in figure~\ref{fig:ml_pipeline}. The final labelled training pulses were passed to the CNN model, which mapped and learnt features in the pulses based on their label. Then the performance of the CNN was evaluated (accuracy, precision, recall, and CV score), where the trained model was passed only the pulses and not the labels, so that the predictions could be made by the CNN and then compared to the actual classifications. Then, the model was passed the pulses without labels from the 20\% test set, and its predictions were evaluated in the same way. The model had never seen the pulses of the test set, and so if the performance evaluations in classifying the two sets (training and testing) are similar, then the model can be seen as generalisable to new data and not likely to overfit. 

\subsubsection{Live data process}
Once the CNN model was confirmed to be performing well, it was incorporated into the digital Compton suppression flow process (figure~\ref{fig:MLCSA}) as the MLCSA step. The processed pulses for a new, live data set, containing pulses from Am-241 and Co-60, were passed to the MLCSA, where the output was pulses with predicted labels of photopeaks (1) or scatter (0).

\subsection{Spectrum} 
The final step in the digital Compton suppression procedure, shown in figure~\ref{fig:MLCSA}, is the production of a spectrum using the MLCSA's predictions, where pulses predicted and labelled as photopeaks (1) were extracted and plotted as a final `peak only' spectrum and the scatter pulses (labelled 0) were discarded. 

\section{Results}\label{sec3}

\subsection{MLCSA performance on the test data set}\label{trained_cnn} 
The CNN model, which was trained on the 80\% data set, was evaluated with the 20\% subset of data. The evaluation results for the test data are shown in table~\ref{tab:ml_training_eval}, which shows that the CNN performed strongly, with all evaluation metrics greater than 70\%. The high CV score indicated that the model is generalisable and not likely to over-fit on new data. 

\begin{table}[h]
\caption{Evaluation results of the trained CNN (trained with data from 5 nuclides) on the 20\% test data set, all values given as a percentage.}\label{tab:ml_training_eval}%
\begin{tabular}{@{}llll@{}}
\toprule
Accuracy & Precision & Recall & CV score\\
\midrule
78.7 & 72.5 & 92.4 & 84.5  \\
\botrule
\end{tabular}
\end{table}

\subsection{MLCSA performance on the live data set} 
\begin{figure}[H]
    \centering
    \includegraphics[width=0.4\textwidth]{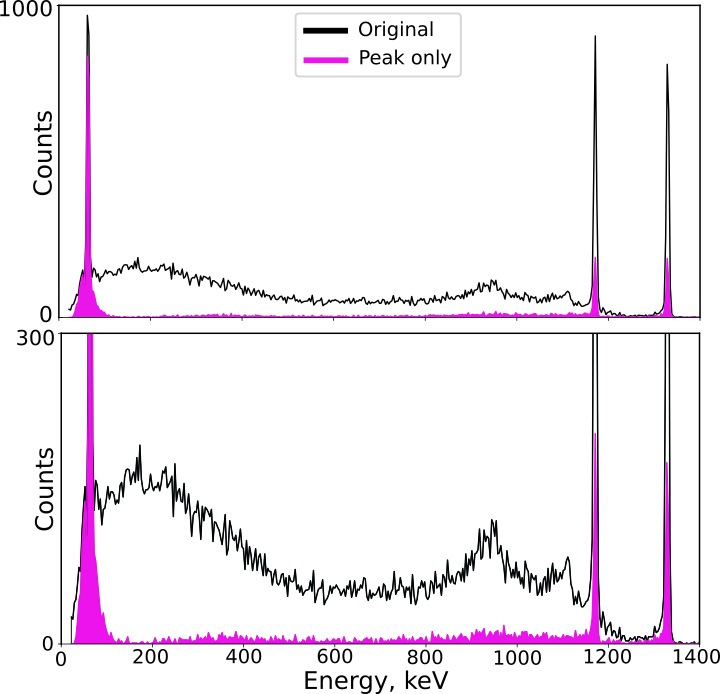}
    \caption{Mixed Am-241 and Co-60 spectrum before (line) and after (filled) the MLCSA. Top shows the full spectrum and bottom is zoomed along the vertical axis}
    \label{fig:MLCSA_final_spec}
\end{figure}

Once the CNN model was trained and evaluated on the small 20\% data set, it was used in the MLCSA with the live data set, comprising an Am-241 and Co-60 source simultaneously positioned 10~cm away from the detector. The data were collected and processed through the digital Compton suppression flow from figure~\ref{fig:MLCSA}, which included the second pulse processing pipeline without the greyed out processes from figure~\ref{fig:ml_pipeline}. The result of the MLCSA is shown in figure~\ref{fig:MLCSA_final_spec}, which shows the before and after spectra.

\subsubsection{Spectrum evaluation metrics}\label{evaluation_metrics}
Three metrics were used to quantify and evaluate the Compton suppression performance of the MLCSA on the final spectrum.

The first was MDA improvement of the Am-241 photopeak. The MDA is a calculation of the lowest measurable activity for a specific set up, and was calculated using the Currie method \cite{currie,MDA_Main} as

    \begin{equation} \label{eq1}   
     \textrm{MDA (Bq)} = \left ( \frac{2.71 + 4.65 \sqrt{B}}{br \times \varepsilon \times t} \right ),
    \end{equation}
\noindent where $B$ is the background sum including a region of 10~keV either side of the photopeak, $br$ is the branching ratio, $\varepsilon$ is efficiency, and $t$ is count time. 

The second metric was signal to noise ratio (SNR). The signal is defined as the net area of the photopeak above the background level and the noise is defined as the area below the peak \cite{net_Area}:
    \begin{equation} \label{eq2}
      \textrm{SNR} = \left (\frac{an}{bg} \right )
    \end{equation}
    where
    \begin{equation} \label{eq3}
     an = \left ( af - bg \right )
    \end{equation}
    and
    \begin{equation} \label{eq4}
      bg = \left ( \frac{C}{2} \right ) \times \left ( B1 + B2 \right )
    \end{equation}
where $an$ is net peak area, $bg$ is the background area underneath the photopeak, $af$ is the full area, C is the number of channels in the photopeak, $B1$ and $B2$ are the numbers of counts at the start and end of the photopeak, respectively \cite{net_Area}.

The final metric was the percentage difference between before and after the MLCSA was applied, including: decrease in MDA of the Am-241 photopeak, reduction in photopeak counts, reduction in max Compton counts on the spectrum for a range of energies, and SNR improvement of Am-241. These were calculated as  
    \begin{equation} \label{eq5}
      \% \textrm{Diff} = \left ( \frac{a - b}{b} \right )  \times 100,
    \end{equation}
\noindent where $b$ and $a$ refer to the relevant value before and after the MLCSA, respectively.

\subsubsection{Spectrum evaluation performance}\label{eval_performance}
 The MLCSA was able to significantly reduce the Compton continuum throughout the whole spectrum, with the largest reductions at the lower energy region (the 200~keV region was reduced by $\sim$~90\%), while almost fully preserving the Am-241 photopeak (the 59~keV photopeak was only reduced by 15\%). The Co-60 photopeaks were preserved, but were significantly reduced in counts ($\sim$~78\%). However, it should be noted that the lower energy (1172~keV) Co-60 photopeak was not provided in the training data set, and so the presence of the photopeak after the MLCSA suggests the CNN model is able to generalise to unseen photopeak pulses. 
 
The percentage reduction in photopeaks in the before and after spectrum was calculated from the data shown in figure~\ref{fig:MLCSA_final_spec}. The reductions for the 59~keV Am-241 photopeak, 1173~keV / 1332~keV Co-60 photopeaks, and the overall counts reduction in two scatter areas, defined as a low scatter region at 200~keV and a high scatter region at 400~keV, are shown in table~\ref{tab:MLCSA_before_After_eval}. The percentage difference was significant on the scatter regions after the MLCSA, while remaining desirably low for the low energy Am-241 photopeak. The Co-60 peaks were reduced significantly, however this is similar to the physical Compton veto system as discussed in Section~\ref{sec:intro} \cite{callums_css} and is an improvement on the output of other digital methods \cite{Tree2019} which completely remove the higher energy photopeaks. 

\begin{table}[h]
\caption{Evaluation of the photopeak counts reduction, and Compton counts reduction (at 200 and 400~keV), as a percentage difference between before and after the application of the MLCSA.}\label{tab:MLCSA_before_After_eval}%
\begin{tabular}{@{}ll@{}}
\toprule
Energy, keV & Peak reduction, \% \\
\midrule
 59 & 15  \\
1173 & 79 \\
1332 & 77   \\
200 & 99 \\
400 & 95 \\
\botrule
\end{tabular}
\end{table}

The SNR was calculated on the Am-241 photopeak using Equations \ref{eq2} - \ref{eq4} for the before and after spectrum, producing values of 1.23 and 1.83 respectively. This produced a significant improvement in the SNR of the Am-24 photopeak, with a percent increase of 49\% (calculated using equation \ref{eq5}), which indicates a reduction in the Compton background at the 59~keV photopeak. 

The MDA of the Am-241 59~keV photopeak was calculated using Equation \ref{eq1} for the before and after MLCSA spectra, and the improvement was calculated using Equation \ref{eq5}. The MDA before was 156~Bq and after was 76~Bq, this produced a reduction of 51\%. Compared to other digital methods that achieved an MDA reduction of 37\% \cite{Tree2019}, this is a significant improvement. 

\section{Discussion}\label{sec4}

The MLCSA reduced the Compton continuum and improved the SNR ratio of the low energy Am-241 photopeak in the presence of higher energy $\gamma$ emitters. The resulting spectrum from the MLCSA on the Am-241 and Co-60 spectrum show that the Compton continuum from Co-60 scatter can be reduced by $\sim$~90\% using machine learning (specifically a basic 1D sequential CNN), while retaining the Am-241 and Co-60 peaks. The SNR ratio of the Am-241 peak was improved by 49\%, and the MDA improved by 51\% - this will lead to better detection and more accurate activity quantification, especially in the case where the lower energy photopeaks are present in lower quantities. This shows the potential for improving Compton suppression systems as this method is similar in performance to the physical Compton veto ring, but with a digital advantage that it could be readily adapted to other detectors.  

This work did not involve complex modelling of a detector, but instead relied on measuring a set of reference sources. Therefore, this could be applied to other detectors as the CNN model would only need re-training with pulses from that detector. This means the MLCSA is potentially detector agnostic, which could be demonstrated with further work and development. Other future developments of this algorithm could include: making the MLCSA run in live time, including more nuclides to the training and testing sets to make it more widely applicable, and improving the machine learning model performance. 

With the significant improvement to low energy nuclide detection in the form of Compton suppression, the MLCSA has the potential to improve fusion diagnostics by improving the detection of key low energy nuclides. As an example, in foil activation experiments, information about the plasma can be calculated from key lines in the gamma spectrum following irradiation of activation foils in a fusion environment \cite{Lees_paper, callums_css, foils_in_fusion_1, foils_in_fusion_2, foils_in_fusion_3, foils_in_fusion_4,foils_in_fusion_5}. The lines of interest are often low energy, can be low abundance, and are often obscured by the Compton scattering of higher energy activation products. The MLCSA could be applied to such spectra, which would reduce the Compton continuum and improve the analysis and therefore could reduce the errors on the fusion power calculations.

The MLCSA developed in this work has the potential to have a broader impact on the nuclear industry. Gamma spectroscopy is a method used in radioactive waste management and disposal (fission and fusion \cite{fusion_waste}), where disposal costs are calculated per unit of activity \cite{epr2016}. Reduced Compton backgrounds from the MLCSA could make gamma spectroscopy systems more accurate, which could have financial benefits relating to disposal costs.


\backmatter

\section*{Declarations}
\subsection*{Funding}
This work has been part-funded by the EPSRC Energy Programme (grant number EP/W006839/1) and the UK STFC, (grant number ST/V001086/1). To obtain further information on the data and models underlying this paper please contact PublicationsManager@ukaea.uk. For the purpose of open access, the author has applied a Creative Commons Attribution (CC BY) licence to any Author Accepted Manuscript version arising from this submission.

\subsection*{Conflict of interest}
The authors have no competing interests to declare that are relevant to the content of this article.

\subsection*{Data Availability}
All data generated or analysed in this work are available on request.

\bibliography{bibliography}

\end{document}